\documentclass[12pt,preprint]{aastex}

\usepackage{emulateapj5}
\usepackage{onecolfloat}
\usepackage{graphicx}
\usepackage{times}
\usepackage[hyperindex,breaklinks]{hyperref}

\newcommand{\kms}{{\rm km/s}\,}
\newcommand{\feh}{{\rm [Fe/H]}}
\newcommand{\afe}{{\rm [\alpha/Fe]}}
\newcommand{\eps}{\epsilon}
\newcommand{\ah}{H06$\,$}

\def\gsim { \lower .75ex \hbox{$\sim$} \llap{\raise .27ex \hbox{$>$}} }
\def\lsim { \lower .75ex \hbox{$\sim$} \llap{\raise .27ex \hbox{$<$}} }

\newcommand{\cutt}[1]{}

\shorttitle{Chemistry and substructures} 
\shortauthors{A. Helmi et al.}

\begin{document}

\twocolumn[

\title{More pieces of the puzzle: Chemistry and substructures in the
 Galactic thick disk}

\author{Amina Helmi\altaffilmark{1}, Mary Williams\altaffilmark{2}, 
 K.C. Freeman\altaffilmark{3}, J. Bland-Hawthorn\altaffilmark{4}, G. De Silva\altaffilmark{5}}

\begin{abstract}
  We present a study of the chemical abundances of Solar neighbourhood
  stars associated to dynamical structures in the Milky Way's (thick)
  disk. These stars were identified as overdensity in the eccentricity
  range $0.3< \epsilon < 0.5$ in the Copenhagen-Geneva Survey by Helmi
  et al. (2006). We find that the stars with these dynamical
  characteristics do not constitute a homogeneous population. A
  relatively sharp transition in dynamical and chemical properties
  appears to occur at a metallicity of $\feh \sim -0.4$. Stars with
  $\feh > -0.4$ have mostly lower eccentricities, smaller vertical
  velocity dispersions, are $\alpha$-enhanced and define a rather
  narrow sequence in $\afe$ vs $\feh$, clearly distinct from that of
  the thin disk. Stars with $\feh < -0.4$ have a range of
  eccentricities, are hotter vertically, and depict a
  larger spread in $\afe$.  We have also found tentative evidence of
  substructure possibly associated to the disruption of a metal-rich
  star cluster. The differences between these populations of stars is
  also present in e.g. ${\rm [Zn/Fe]}$, ${\rm [Ni/Fe]}$ and ${\rm [SmII/Fe]}$, suggesting a real physical distinction.
  \end{abstract}

]





\altaffiltext{1}{Kapteyn Astronomical Institute, University of Groningen, 
 P.O.Box 800, 9700 AV Groningen, The Netherlands. 
{\sf{e-mail: ahelmi@astro.rug.nl}}}
\altaffiltext{2}{Astrophysikalisches Institut Potsdam, 
 An der Sternwarte 16, D-14482, Germany.
{\sf{e-mail: mary@aip.de}}}
\altaffiltext{3}{Research School of Astronomy \& Astrophysics
The Australian National University, Cotter Road Weston Creek, ACT 2611, Australia}
\altaffiltext{4}{Sydney Institute for Astronomy, School of Physics A28, University of Sydney, NSW 2006, Australia}
\altaffiltext{5}{Australian Astronomical Observatory, PO Box 915, North Ryde, NSW 1670, Australia}

\section{Introduction}
\label{sec:intro}

In the concordance $\Lambda$CDM cosmological model, mergers are 
ubiquitous, and expected to have left imprints in galaxies like the Milky Way
\citep{hw99,bj05}. Although the majority of stars in disks have
probably formed in situ, some fraction may originate in accreted
objects on low inclination orbits \citep[e.g.][]{abadi03a}. Some evidence of such events
may be the Monoceros ring in the outskirts of our Galaxy
\citep[][although its origin is still debated]{newberg}, the
low-latitude substructures in M31 \citep{richardson}, the Arcturus
stream \citep{nhf}, the $\omega$Cen stream 
\citep{Meza,Majewski}, and the groups found by \citet[][hereafter
  H06]{ah} and \citet{AF2006} near the Sun. 

The recovery of such events is the focus of many 
surveys, and correspondingly, attention has been given to the
optimal search techniques. Most of the substructures
discovered thus far have been identified in projections of
phase-space, namely spatially \citep[e.g.][]{vasily-fos},
kinematically \citep{gilmore2002} or in an ``integrals of motion''
space \citep{h99}.  However, there
is the intriguing suggestion that merger debris may also be
identifiable through peculiar chemical abundance patterns
\citep{kcf,desilva}, as stars formed in a common
system must have experienced similar star formation and enrichment
histories. Some first detections of chemically peculiar sequences or clustering have been reported by \citet[][]{NS2010,NS2011}
and \citet{WdB2010, WdB2012}.

In this Paper we present chemical abundances for a sample of nearby
stars from the Copenhagen-Geneva survey found to define an overdensity
in the eccentricity range $0.3 \le \epsilon \le 0.5$ by \ah. We have
performed a high-resolution follow-up study of 72 of these stars, and
have supplemented this dataset with that of \citet[][hereafter
ST]{stonkute2012,stonkute2013} of 21 stars in the same overdensity.
As we shall see below, a few subpopulations of stars with different
dynamics and chemical abundances co-exist in this region of Galactic
(extended) phase-space.

\section{Observations and Analysis}

\subsection{The puzzle stars}

The stars in \ah were identified in the Copenhagen-Geneva survey
\citep[][GCS hereafter]{gcs}, because of their peculiar distribution in the space of
orbital apocentre-pericentre and $z$-component of the angular
momentum. These stars share similar relatively planar orbits with
moderate eccentricities between 0.3 and 0.5. This eccentricity region
is overdense with respect to what is expected for a smooth model of
the Galaxy (see Figures 10, 11 and 13 of H06).  Besides their
characteristic dynamics, the 274 stars identified were found to have
distinct metallicity and age distributions. A separation into three
different Groups was proposed on the basis of their metallicity,
namely G1: $-0.45 \!<\!  \feh \!<\!  -0.2$~dex, G2: $\langle \feh
\rangle \!\sim\!  -0.6$~dex, and G3: $\feh \!<\!  -0.7$~dex, where as
the metallicity decreases, there is a tendency for the groups to have
larger mean eccentricity. Group G2 has some overlap with the Arcturus stream, in terms
of its average $V$ velocity and mean $\feh$. Note that these $\feh$ were estimated from
Str\"omgren photometry and are slightly different from those derived here.

\begin{figure}[t]
\begin{center}
\includegraphics[width=83mm]{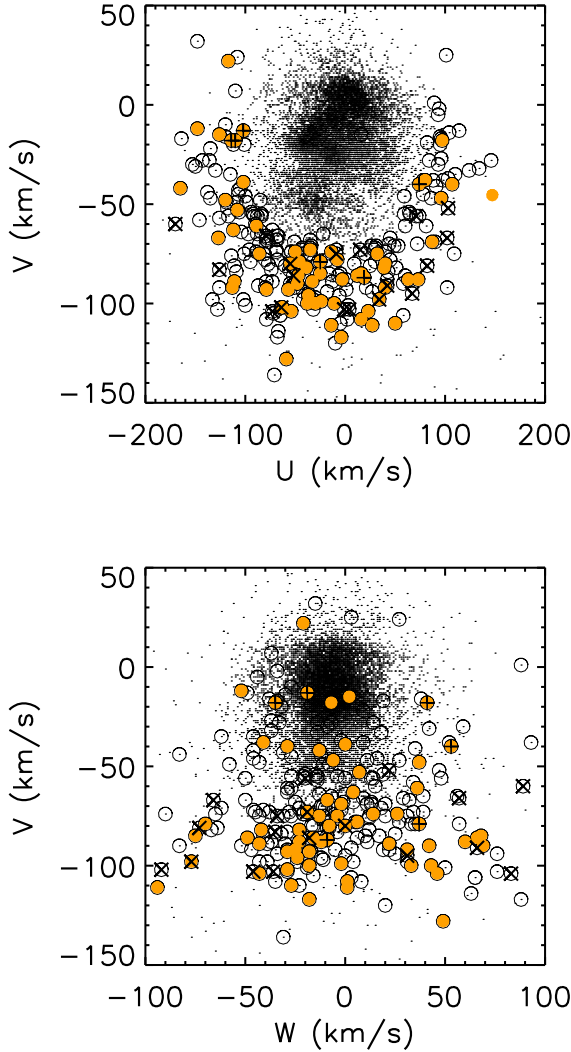}
\end{center}
\caption{Velocities of stars in the revised GCS catalog, highlighting
those from \ah. The solid symbols correspond to those
we followed up with high-resolution spectroscopy, while the crosses are
from ST. We have kept only stars with revised eccentricities
$0.275 \le \eps \le 0.525$. The stars with ``+'' symbols are those in the ``tentative'' cluster (see main text).
\label{fig:velocity-puzzle}}
\end{figure}

Figure \ref{fig:velocity-puzzle} shows the velocity distribution of
the revised GCS \citep{holmberg2009} highlighting the \ah stars. The stars included in this {\it Paper} are indicated with solid 
symbols (those we followed up with high-resolution spectroscopy), or with crosses (those from ST).

\subsection{Derivation of chemical abundances}

We performed our high signal-to-noise high-resolution spectroscopic
study using
UCLES on the 4m Anglo Australian Telescope. It includes 
36 members of G1 (27.5\% of all candidates identified by
\ah), 22 members of G2 (25.6\%), and 14 members of G3 (20.6\%). 
The wavelength coverage for our observations was $3850$ -
$5300\textrm{\AA}$ and at least three exposures of on average
$400\ \textrm{s}$ were done per star at a resolution $R \sim
45,000$. A signal-to-noise of $\sim100-150$ per pixel at the central
wavelength was obtained. The data were reduced using the standard IRAF
routines \textit{ccdred} and \textit{echelle}, which included bias
subtraction, flat-fielding, order extraction, scattered light
subtraction and wavelength calibration. We focus here on those stars with $(V-K) > 1.4$. Furthermore, two
stars in G1 have significantly different eccentricities in the revised GCS, so we do not consider these further. 
We are thus left with a total of 64 stars: 30 from G1, 21 from G2 and 13 from G3.

To supplement this sample we also consider the 21 stars originally in G3 followed up by ST.
Two of these have also been observed by us, and 3 have significantly revised GCS eccentricities which place them outside of the region of interest. We are therefore left with 16 independent stars from this sample.

Elemental abundances were derived for our stars by performing a Local
Thermodynamic Equilibrium (LTE) analysis with the MOOG code
\citep{Sneden1973}. To measure equivalent widths we used a modified
version of the DAOSPEC program which automatically fits Gaussian
profiles to lines in a spectrum \citep{PS2008}. The modifications involved disabling DAOSPEC's continuum
fitting, and instead performing hand-fitting for our crowded blue
data. 
In this Paper we present the abundances for our
stars using stellar parameters calculated using a `spectroscopic' (rather than physical) approach
 as they yielded better agreement for stars in
common with external studies by \citet{Reddy2003,Reddy2006} and
\citet{Bensby2003,Bensby2005}. In this approach we force excitation
and ionisation balance to derive $T_{\rm{eff}}$ and $\log g$
respectively. Microturbulence is found by requiring that there is no
dependence of abundance on line strength. We present here abundance
results for Mg, Ca, Ti, Cr, Ni, Zn, Nd and Sm, using elements or lines
for which hyperfine and isotopic splitting need not be considered. The
first three elements represent the $\alpha$-elements (where
$[\alpha/\textrm{Fe}]=[\textrm{(Mg+Ca+Ti)}/3\textrm{Fe}]$), Cr, Ni, Zn
represent Fe-peak elements, and Nd and Sm have a varying mix
of $s$- and $r$-process contributions \citep{Arlandini1999, Burris2000}.

\subsection{Results}

We take here a fresh look at the H06 stars now supplemented by the
chemical abundances, rather than follow the originally proposed
separation into groups.

\begin{figure}
\includegraphics[width=83mm]{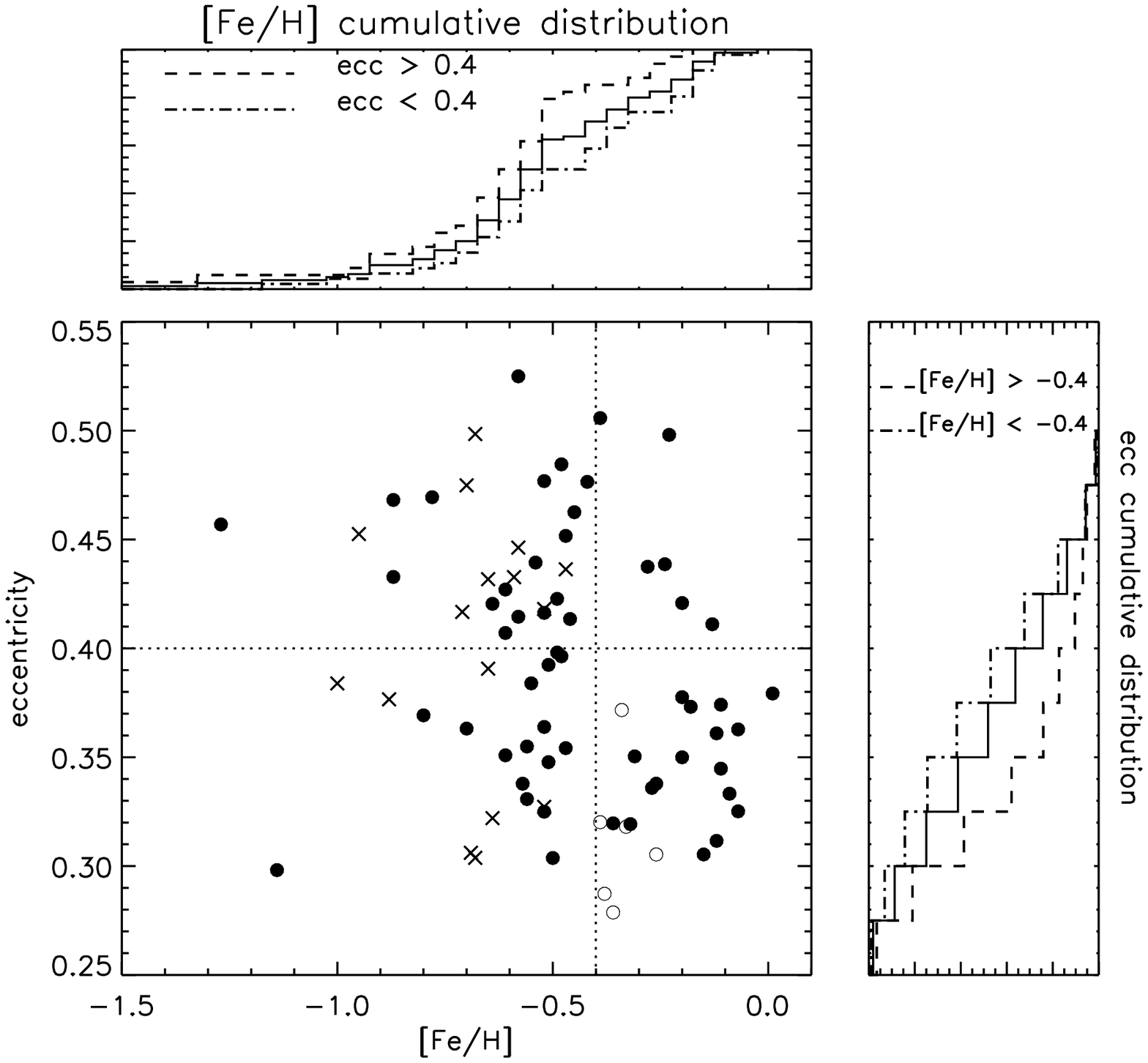} \\
\includegraphics[width=60mm]{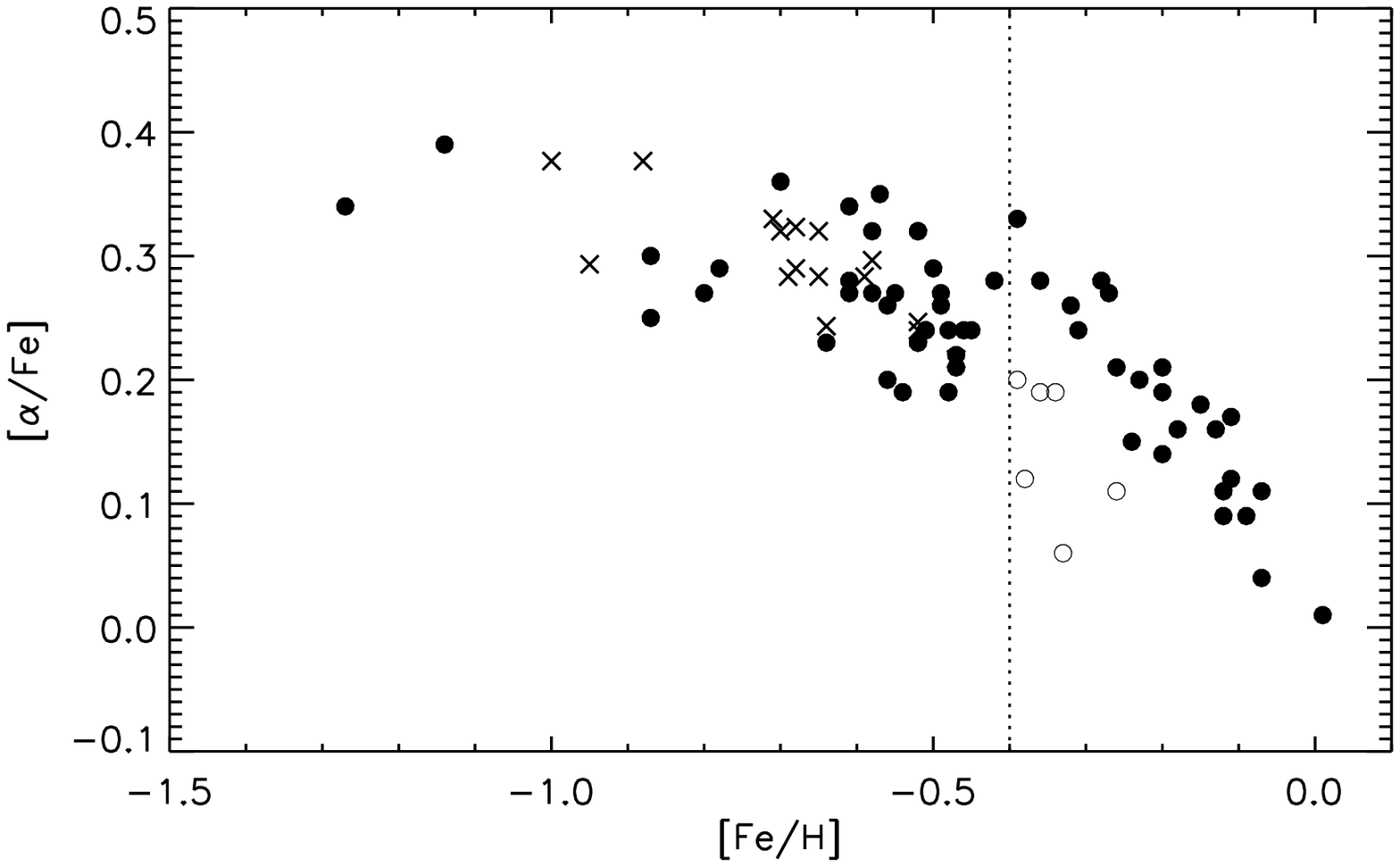}
\caption{Eccentricity vs $\feh$ (central panel), $\afe$ vs $\feh$ (bottom panel), and
1D cumulative distributions of $\feh$ (top) and eccentricity (right)
  for stars in our sample (solid black) and those in ST (crosses). 
In the bottom panel we have indicated separately with open circles  those stars possibly 
associated to a disrupted cluster. The lines at $\feh = -0.4$ and $\eps = 0.4$
indicate where a transition in populations seems to be present in our dataset. 
\label{fig:alpha}}
\end{figure}

The central panel of Figure \ref{fig:alpha} shows the distribution of
orbital eccentricity $\epsilon$ for the H06 stars as function of
$\feh$. Note that there appears to be a lack of stars in the upper
right quadrant, i.e. for $\epsilon > 0.4$ at $\feh > -0.4$ (also the
case in the full H06 sample). This is also evidenced in the uppermost panel
where we compare the $\feh$ cumulative distribution for the full sample
(solid), for stars with $\epsilon > 0.4$ (dashed), and for those
with $\epsilon < 0.4$ (dotted-dashed). The probability that the latter
two are drawn from the same parent metallicity distribution is
$1.1\times 10^{-2}$ according to a KS test. On the other hand, the
cumulative distribution of $\epsilon$ is plotted in rightmost panel of the
figure. The probability that the stars with $\feh < -0.4$
(dotted-dashed) and those with $\feh > -0.4$~dex (dashed) have the same eccentricity distribution is $4.7\times
10^{-4}$ according to a KS test.  Also intriguing is that the stars
with $\feh <-0.4$~dex have on average a larger vertical velocity
dispersion ($47.7$ \kms) compared to the group with high metallicity
($\sigma_z \sim 31.4$~\kms).
\begin{figure}
\centering 
\includegraphics[width=80mm]{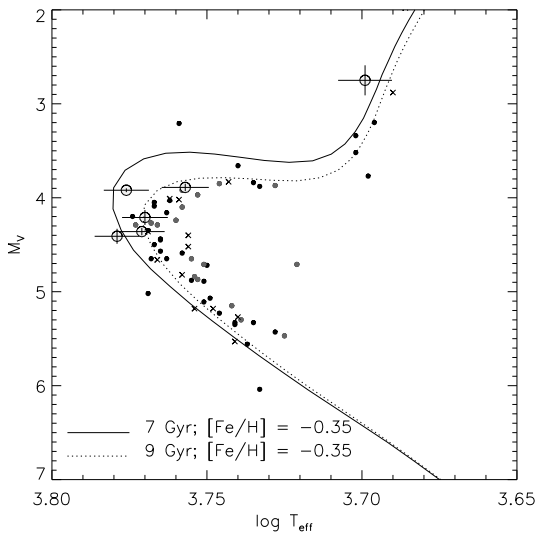} 
\caption{HR diagram for the stars that we followed up spectroscopically (solid symbols), 
and those from ST (crosses). Those in the tentative cluster identified in the bottom panel of Fig.~\ref{fig:alpha}
are indicated with larger light grey symbols with error bars. Stars with [Fe/H] $> -0.4$ dex are shown in dark grey.
The temperature error bars correspond to $\pm 100$~K. The curves plotted 
are Yonsei-Yale isochrones
\citep{yy} with $\feh = -0.35$ and $\afe = 0.2$.
\label{fig:cmd_cluster}}
\end{figure}

The bottom panel of Figure \ref{fig:alpha} shows $\afe$ vs $\feh$ for
the stars in our sample, where the black solid symbols indicate those
that we have followed up and the crosses are from ST. We have
highlighted with open symbols a set of 6 stars with $\feh \sim -0.35$
dex. Closer examination of these stars reveals that 3 of them define a
very cold kinematic group (visible in Fig.~\ref{fig:velocity-puzzle}
at $V \sim -15$ and $U \sim -100$ \kms, + symbols). These stars' eccentricities
are 0.28, 0.31 and 0.32 while their $\feh = -0.36$, $-0.26$, and
$-0.33$ dex. The $\eps$ and $\feh$ of the remaining 3 stars are also
similar as shown in the central panel of Figure \ref{fig:alpha}. 
This is also true for the chemical abundances of these stars (Fig.~\ref{fig:all_feh} below), 
for all measured elements except for [Mg/Fe] (where even those stars in the kinematic group depict a larger spread).  

There are two possible interpretations that we may put forward: $i$)
that the other 3 stars with similar $\feh$ are a random sample of the thin disk, whose
orbits are slightly hotter than typical for this component; $ii$) that
they are part of the ``kinematic clump'' of 3 stars (a disrupted cluster?) because they are
not very different dynamically, they have similar $\feh$ and other
abundances, as well as being indistinguishable in age (and clearly
younger than the upper $\afe$ sequence at this
metallicity). This is evident from Figure \ref{fig:cmd_cluster}, where
we plot the HR diagram of the stars in our sample. The lines
correspond to 7 (solid) and 9 (dotted) Gyr old Yonsei-Yale isochrones
\citep{yy} with $\feh = -0.35$ and $\afe = 0.2$.  We have
also used a lighter gray colour for the stars with $\feh > -0.4$ dex
but higher $\afe$. This allows comparing the age of the stars
in the tentative cluster to those stars with similar metallicity and
eccentricity. Clearly those ``cluster'' stars are on average younger
(as they have bluer colours, higher temperatures), and consistent with
being a single age population. This motivates us to have a slight preference for the
``cluster'' scenario as it is not easy to understand why the
contamination from the thin disk (which could be present for
eccentricities $\sim 0.3$) would preferentially select stars with such
a narrow range of $\feh$. Our sample has no bias against stars with
high metallicities, so one would perhaps have expected contaminants
from a broader range of $\feh$ more representative of the whole of the
thin disk.

In summary, the bottom panel of Figure \ref{fig:alpha} shows that at high $\feh$,
stars define a narrow sequence depicting a steep relation between
$\afe$ and $\feh$, and that a transition occurs at $\feh\sim
-0.4$ dex, below which the scatter in $\afe$ increases
significantly. Therefore the joint analysis of the panels in
Fig.~\ref{fig:alpha} suggests that at least two, but likely three,
populations co-exist in our dataset: a low eccentricity population that dominates at high $\feh$, a
high eccentricity population that is essentially only present at low $\feh$, and possibly a disrupted cluster. Note
that none of these first two populations could belong to the thin disk
given their kinematics and abundance trends. This is in agreement with the recent 
analysis of the Gaia-ESO survey by \cite{ale} that also shows a dearth of stars with high eccentricities at high $\feh$ in the
thick disk.

Figure \ref{fig:all_feh} shows various element ratios as function of
$\feh$ for all the stars in our sample. We have used the same symbol
scheme as in the bottom panel of Fig.~\ref{fig:alpha}, while the
colour coding (from dark to light) and symbol size (from small to
large) indicate increasing values of the eccentricity. The error
bars correspond to two different estimates of our measurement
uncertainties: from repeat observations of the same star (smaller bar)
and from a global estimate\footnote{which is likely somewhat overestimated given the results
from repeats.} (larger bar) derived from the change in the
abundance of a given element as $T_{\rm eff}$ varies by $\sim \pm
100$~K, log~$g$ by $\pm 0.1$ and $\feh$ by $\pm 0.1$.  The
black points in the panels correspond to stars from
\citet{Reddy2003,Reddy2006} that have $\epsilon < 0.3$ or $\epsilon >
0.5$ (to provide a comparison sample with different kinematical
characteristics from the H06 set). 
To be on the same scale, and using
15 stars we have in common, we applied a zero-point correction $\Delta
= {\rm [El/H]}_{us} - {\rm [El/H]}_{Reddy}$ to these
abundances, namely $\Delta {\rm [Fe/H]} = 0.09$, $\Delta {\rm [Mg/Fe]} = 0.06$,
$\Delta {\rm [Ca/Fe]} = 0.08$, $\Delta {\rm [Ti/Fe]} = 0.01$, $\Delta {\rm [Zn/Fe]} =
0.02$, $\Delta {\rm [Cr/Fe]} = -0.03$ and $\Delta {\rm [Ni/Fe]} = -0.03$
\citep[for more details, see][]{Williams-thesis}. The stars in ST have been put first on the Reddy
scale, and later shifted in the same way. The dashed lines in the [Mg/Fe] vs $\feh$ panel correspond to the sequences
for the thin and thick disks as determined by \citet[][]{fuhrmann2011} in his volume complete
sample of nearby stars.

The three top left panels of this figure show that stars with $\feh >
-0.4$ define a relatively narrow track in [Mg/Fe], [Ca/Fe] and
[Ti/Fe], with an internal dispersion smaller than the global estimated
error, or even than the error determined from repeats. Note also that
[Zn/Fe] and, even [Ni/Fe] follow clearly distinct tracks as function
of $\feh$, despite the fact that both Zn and Ni are Fe-peak
elements. Recall that in this region of chemical space, most stars
have lower eccentricities (darker symbols). We see once
more that a transition in the chemical properties of the stars appears
to occur at $\feh \sim -0.4$, as the scatter substantially increases
for all elements below this metallicity \citep[also in ${\rm [SmII/Fe]}$,
which is reasonable if this is largely an $r$-process product of
massive stars,][]{Mish2013}.  Also interesting is that the sequence
delineated by the stars with $\feh > -0.4$ appears to define an upper
envelope at lower $\feh$ (as there is a tendency for the darker
symbols to have higher $\afe$ at these lower metallicities, and
perhaps also lower [SmII/Fe]). The higher eccentricity and more
metal-poor stars (lighter colours and larger symbols) follow well the
sequence of ``thick'' disk stars, and given their kinematics, could be
part of the canonical thick disk (as suggested by e.g.\ the uppermost
dashed line in the [Mg/Fe] vs $\feh$ panel).
\begin{figure*} 
\centering
\includegraphics[width=80mm]{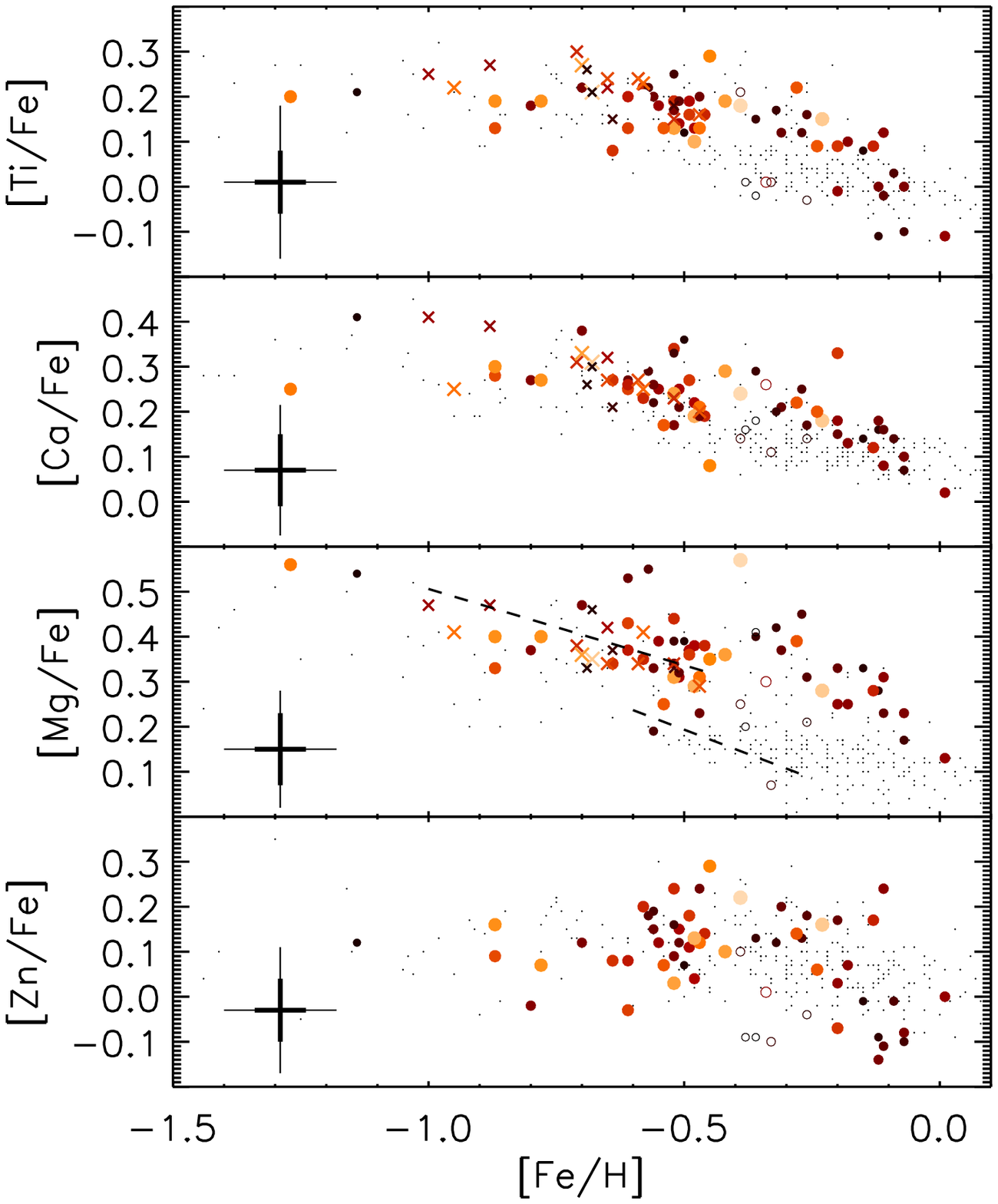}
\includegraphics[width=80mm]{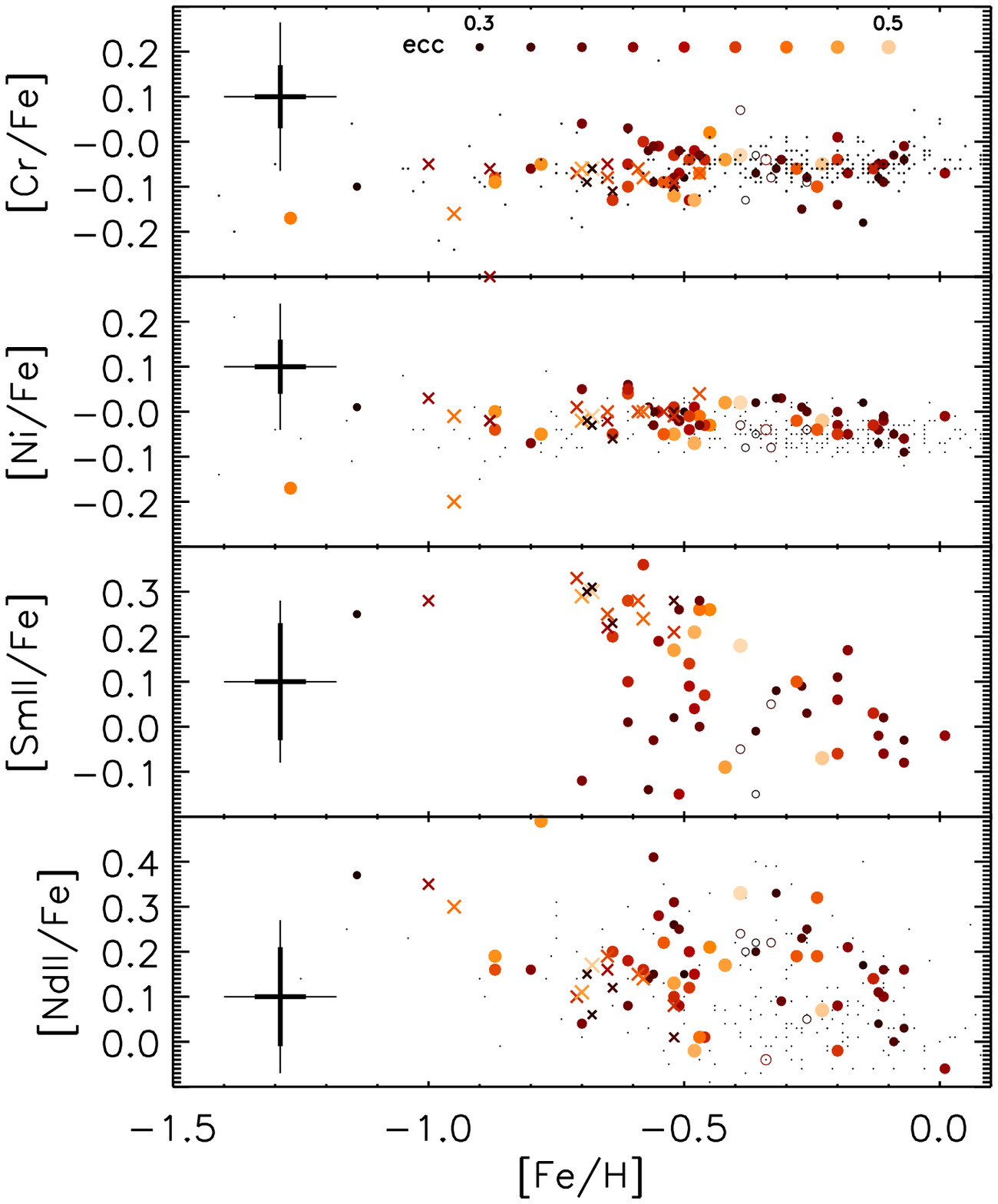}
\caption{Various elements vs $\feh$ for our sample of H06 stars (circles, where open symbols are for the tentative cluster) and for ST (crosses). The colour-coding (dark to light) and symbol size (small to large) indicates increasing eccentricity as shown in the top right. For comparison, we have also plotted (small dots) stars with $\epsilon <
  0.3$ and $\epsilon > 0.5$ in the thin and thick disks from
  \citet{Reddy2003,Reddy2006}. The error
  bars indicate two estimates of our uncertainties: from repeats and from varying the derived
  astrophysical parameters.
\label{fig:all_feh}}
\end{figure*} 

Therefore the stars with lower $\eps$ tend to define the upper envelope of the $\afe$
vs. $\feh$ both in our joint sample as well as in Reddy's. It
is also interesting to note that the highest $\feh$ stars have lower
[Zn/Fe], [Ti/Fe] than even the thin disk stars at this metallicity.  On the other
hand, their [Ca/Fe] is similar while their [Mg/Fe] is higher even than
the thick disk stars at these $\feh$.

The stars indicated with open symbols define very tight structures
in [Ni/Fe] and also in [Ti/Fe]. In the other elements the scatter is
consistent with the estimated global error. We may thus tentatively conclude that the 
indication of a single age (Fig.~\ref{fig:cmd_cluster}), 
their location in abundance space (which overlaps generally with thin disk stars), and their tight kinematics
all are consistent with a (star) cluster origin for these stars. 


The two stars with $\feh \sim -0.87$ (HD 25704 and HD 63598) are particularly intriguing,
since they have nearly identical compositions in all elements. They
also share the same location in the HR diagram, implying that they
have the same age \citep[also confirmed by][]{casa2011}. Furthermore, their distances are essentially
identical ($\sim 50$ pc). According to their kinematics, they would be
on relatively eccentric orbits, with $\eps = 0.43$ and $0.47$ and
velocities ($U$, $V$, $W$) = ($-127$, $-67$, $-9$) km/s and
($16$, $-108$, $1$) km/s, with average errors of 3 and 2
km/s respectively. A possibility is that these stars constitute a 
binary, as their velocity difference is too large for them to be
located on the same orbital phase of a tidal stream.

\section{Discussion}

We have shown that stars with $0.3 \le \epsilon \le 0.5$ do not
form a homogeneous population. In this eccentricity range, a few
populations with different characteristics are found:
\begin{itemize}
\item High metallicity stars ($\feh > -0.4$~dex) predominantly have lower eccentricities ($\epsilon \le 0.4$), cold
  vertical kinematics, are $\afe$ enhanced, and define a relatively
  narrow track in all $\alpha$-elements vs $\feh$, as well as in Zn and
  Ni. While their kinematics might suggest they could constitute the
  tail of the thin disk population, their abundances clearly rule out
  this possibility. This population is dominated by stars with an average age of 8 Gyr, and has no stars older than 12 Gyr.

\item Lower metallicity stars can be on high or low eccentricity orbits, and show a
  larger abundance scatter than the population described in the
  previous item, and are also on average older, i.e.\ $\sim 9$ Gyr, with several stars as old as 14 Gyr. The higher eccentricity stars 
 could be considered representative of the canonical thick disk. 

\item The third group of stars present in our dataset have very
  similar iron-peak abundances, and a somewhat larger scatter in $\alpha$
  elements, especially in [Mg/Fe]. Half of these stars are also
  kinematically clustered in a tight clump, and all have similarly low eccentricities and ages, suggesting that they
  form a dynamical entity with a scale perhaps similar to that of a
  (star) cluster.
\end{itemize}

It is interesting to note that the first population overlaps with the
high $\alpha$ metal-rich population in \citet{adi} (who also miss
high-eccentricity metal-rich stars, see their Fig.~5). However in our
sample, as in \citet{bensby2013}, there is no gap in $\feh$ between this population and the more metal-poor stars.

The issue now is how to interpret these two main populations (i.e. excluding the tentative cluster) in the context of
the evolution of our Galaxy. Let us consider the following
possibilities.
\begin{itemize}
\item {\it Mergers:} In the most widely discussed scenario for the
  formation of the thick disk, a pre-existing disk is heated up via a
  minor merger. Such a disk would be more massive and hence have a
  higher metallicity, and lower eccentricity than the accreted
  population \citep{sales}. Thus we would be tempted to identify this
  population with the sequence defined by the more metal-rich stars in
  our dataset, and the intruder with the higher eccentricity and lower
  $\feh$ stars. The colder kinematics of the first group, ages and
  enhanced $\afe$ (i.e. higher SFR indicating more massive) would also
  seem to support this view.

\item {\it Canonical thick disk and bulge:} Stars with higher
  eccentricities could be associated to the canonical thick disk,
  while those more metal-rich, could on the basis of their abundance
  patterns and ages be the extension of the bulge to the Solar
  neighbourhood. If this scenario is enforced, it would seem to be
  necessary to argue that the bulge/bar stars eccentricities and
  vertical motions are on average lower than those of the canonical
  thick disk, which appears
  slightly counterintuitive given our knowledge of in-situ bulge stars
  \citep[e.g.][]{soto}.

\item {\it Radial migration:} In this scenario the high metallicity 
  population would originate from the inner galactic disk
  \citep[][]{grenon,sch}. Hence one might expect such stars to have higher
  eccentricities (although migrated stars can also be on nearly circular orbits) 
  as well as hotter $\sigma_z$ on average as postulated by \cite{sch} \citep[although see][who argue that if the 
vertical action is conserved, then the stars migrating from the inner disk become colder]{minchev}. On the other hand, 
the population of low $\feh$ could not be considered the 
migrated population from the outer disk \citep[as in the model by][]{Haywood}, because its $\sigma_z$
is higher than that of the thin disk locally, contrary to the prediction from simulations including radial migration 
\citep[e.g.][]{solway,minchev}.
\end{itemize}

Which of these three scenarios is more likely is unclear, although the
last one appears less plausible at face value in accounting for all the features observed. 

It is interesting that we have not found that stars cluster in tight
lumps at a fixed metallicity, except in a few occasions. Rather we
have found sequences as expected for populations that have had a
finite time to evolve (and self-enrich). The chemical evolution tracks
are probably indicative of fairly intense star formation histories rather unlike those of e.g. dSph.

Our work suggests that the Milky Way's disc(s) are the overlap of
several populations with distinct characteristics. The combination of
dynamics, stellar population analysis and chemical abundances have
allowed us to establish this. However, we seem to be still a long way
away from being able to deconstruct and understand the history of the
Milky Way.

\acknowledgements

We thank Peter Stetson for his DAOSPEC code used in the equivalent
width analysis. We thank Jon Fulbright for providing a
model-atmosphere interpolation code, with kind permission of Jennifer
Johnson. AH was partially supported by ERC-StG
GALACTICA-240271 and NWO-VIDI grants. This research made use of
the Vienna Atomic Line Database, Austria, the
SIMBAD database at CDS, France and the NASA ADS,
USA.

\end{document}